# AI-Driven Feedback Loops in Digital Technologies: Psychological Impacts on User Behavior and Well-Being

Anthonette Adanyin
A.adanyin@wlv.ac.uk
*Department of Data Science and Artificial Intelligence, School of Engineering, University of Wolverhampton, UK.*

**Abstract**
The rapid spread of digital technologies has produced data-driven feedback loops, wearable devices, social media networks, and mobile applications that shape user behavior, motivation, and mental well-being. While these systems encourage self-improvement and the development of healthier habits through real-time feedback, they also create psychological risks such as technostress, addiction, and loss of autonomy. The present study also aims to investigate the positive and negative psychological consequences of feedback mechanisms on users' behaviour and well-being. Employing a descriptive survey method, the study collected data from 200 purposely selected users to assess changes in behaviour, motivation, and mental well-being related to health, social, and lifestyle applications. Results indicate that while feedback mechanisms facilitate goal attainment and social interconnection through streaks and badges, among other components, they also enhance anxiety, mental weariness, and loss of productivity due to actions that are considered feedback-seeking. Furthermore, test subjects reported that their actions are unconsciously shaped by app feedback, often at the expense of personal autonomy, while real-time feedback minimally influences professional or social interactions. The study shows that data-driven feedback loops deliver not only motivational benefits but also psychological challenges. To mitigate these risks, users should establish boundaries regarding their use of technology to prevent burnout and addiction, while developers need to refine feedback mechanisms to reduce cognitive load and foster more inclusive participation. Future research should focus on designing feedback mechanisms that promote well-being without compromising individual freedom or increasing social comparison.

## 1.0 Background/Introduction

The rapid expansion of digital technology has introduced data-driven feedback loops into everyday life, reshaping behaviors and mental states in many ways. Wearables, social media platforms, and mobile apps have become deeply integrated into daily routines, generating continuous streams of personalized feedback. These feedback systems often influence behavior subtly, leaving users unaware of the extent to which their habits and emotions are shaped by them. Feedback mechanisms built into digital technologies offer several benefits. Wearables, such as fitness trackers and smartwatches, are designed to promote healthier lifestyles by providing users with real-time data on their physical activities, sleep patterns, and other health-related metrics. Studies suggest that such devices can motivate behavior change by fostering self-improvement and creating new habits. Competition and cooperation features integrated into wearables promote user engagement, with competitive settings driving performance and cooperative ones enhancing satisfaction and well-being (Wolf et al., 2020). Additionally, personalized feedback on sleep, provided through mobile apps, has been found to improve sleep consistency and physical symptoms like fatigue and stiffness, which enhance users' overall well-being (Takeuchi et al., 2022).

Mobile apps have also become an essential tool for mental health support. Digital interventions using mobile platforms have shown promising results in improving mental health outcomes, particularly for individuals experiencing high levels of stress. According to Collins et al. (2020), mindfulness apps deployed in workplace settings have been found to reduce anxiety and depressive symptoms while improving resilience and stress management skills (Collins et al., 2020). Moreover, apps that use gamification elements such as points, challenges, or social features have proven effective in fostering continuous engagement, further demonstrating the value of feedback systems in promoting mental well-being and self-improvement (Kaur et al., 2023).

However, feedback loops also present significant psychological risks. Social media platforms utilize feedback mechanisms to keep users engaged through features such as likes, comments, and notifications. While these interactions create a sense of connection, they also expose users to constant social comparison. Adolescents who receive fewer likes on social media platforms report feelings of rejection and negative self-evaluation, contributing to increased levels of emotional distress and depressive symptoms (Lee et al., 2020). Similarly, cognitive overload from continuous exposure to social media content has been linked to elevated stress and reduced well-being, especially in older users (Matthes et al., 2020). Another negative aspect of feedback loops lies in their potential to foster addictive behaviors. Social media platforms and mobile apps use dopamine-driven reward systems to encourage users to engage with their content repeatedly. This creates habits of frequent checking, with the reward systems reinforcing compulsive behaviors, often to the detriment of mental health. Research has shown that habitual social media checking behaviors are associated with lower neural sensitivity to social rewards and can alter the brain's development, especially during adolescence (Maza et al., 2023). This addictive potential highlights the need to balance the positive social connections facilitated by these platforms with the risks of compulsive usage. Media dependency also emerges as a critical issue in the context of feedback loops. Many users rely on mobile social apps not only for communication but also as a source of validation and emotional regulation. When individuals develop dependencies on these systems, it impairs their ability to self-regulate and exacerbates problematic behaviors (Liu et al., 2020). Furthermore, the design of feedback systems is often optimized for engagement rather than user well-being. Social media platforms and mobile apps are engineered to retain users' attention, which can exacerbate mental health problems. Feedback that emphasizes competition may increase performance among certain users, but it can also intensify fear of failure, undermining psychological well-being in the long term (Wolf et al., 2020). Additionally, social comparison mechanisms embedded in these platforms perpetuate unrealistic standards, contributing to dissatisfaction and emotional strain (Lee et al., 2020).

The integration of feedback loops into digital technologies brings both benefits and challenges to users' mental states and behaviors. While wearables and mobile apps can motivate positive behavior changes and enhance mental well-being, the addictive nature of feedback systems and the psychological stress induced by social media raise concerns. As digital technologies continue to evolve, it is essential to develop strategies that maximize their positive impacts while minimizing the psychological risks they pose.

## 1.1 Aim

This study aims to explore the psychological impacts of data-driven feedback loops. Specifically, it examines how wearables, social media, and mobile apps influence user behavior, motivation, well-being, and unintended psychological effects, both positive and negative.

## 2.0 Literature Review

## 2.1 Conceptual Review

The wide diffusion of data-driven feedback mechanisms, particularly in wearable devices, social media, and mobile apps, significantly revolutionizes the way human conduct and psychological processes take place. These technologies make data available instantly to alter choices made, social exchanges, and mental health in ways that might be advantageous but also possibly dangerous.

### (a) Positive Impacts of Data-Driven Feedback Loops

Wearable devices and self-monitoring applications enhance positive behavioral change through heightened self-awareness and accountability. Health and fitness wearables are designed to provide immediate feedback on physical activities, thereby associating with improved physical health and well-being (Baker 2017). Individuals using such technologies report increased motivation and goal attainment in continuous tracking of progress. Equally, social media allows users to create social capital through bonding and bridging social interactions. This cultivation of belongingness and interconnectedness contributes favorably to psychological well-being (Ostic et al., 2021). In both professional and social contexts, feedback mechanisms facilitated by wearable technologies have demonstrated potential in improving communication and performance. In a study by Damian et al. (2015), feedback given to presenters regarding their posture and vocal tone during presentations enhances their delivery and self-assurance, thereby establishing a cycle of positive reinforcement. Furthermore, the incorporation of haptic feedback via wearable devices has the potential to enhance task performance by augmenting the user's perception of social presence and engagement (Hadi & Valenzuela, 2020). When these feedback, systems are integrated into digital health interventions, they facilitate favorable behavior modifications, including heightened levels of physical activity and improved lifestyle decisions (Zhao et al., 2016). Research suggests that both bonding and bridging forms of social capital account for 45.1% of the variation in well-being linked to social media usage (Ostic et al., 2021). Moreover, haptic feedback provided by smartwatches contributes to an increase in users' engagement with tasks by fostering an enhanced perception of social presence (Hadi & Valenzuela, 2020).

### (b) Negative Psychological Effects and Technostress

Although feedback loops stimulate proper behavior, there are also various serious negative psychological impacts. One critical issue is the emergence of technostress, developed by users who feel overwhelmed by continuous notifications and feedback. The concern is more widespread among individuals with excessive social media use, where there is a link with higher levels of stress, mental fatigue, and lower levels of productivity (Hsiao et al., 2017). Research highlights that 324 surveyed professionals reported heightened technostress due to constant exposure to social media notifications, which directly contributed to workplace cyber incivility (Mert et al., 2021). Similarly, compulsive use of mobile social apps is positively linked to anxiety and stress among students, contributing to a decline in academic performance (Hsaio et al., 2017). Similarly, wearable devices generating immediate biometric feedback can also make users dependent on them and focus on data-driven activities rather than internal drives (Baker, 2017). Another negative consequence relates to the concept of social comparison. Individuals on social media also engage in frequent comparative evaluations and seeking behaviors that often lead to the emergence of anxiety and depression symptoms, especially among teenagers (Nesi & Prinstein, 2015). Adolescents exhibiting fear of missing out (FOMO) exhibit compulsive behavior linked to increased levels of stress and decreased life satisfaction (Oberst et al., 2017). This type of behavioral addiction undermines individual freedom as such individuals become increasingly dependent upon feedback mechanisms for emotional validation and social acceptance. Excessive reliance on smartphone dependency has also been encouraged. Studies have shown that by monitoring students' behaviour such as screen time and locomotion patterns mental health could be predicted with an accuracy rate of 79% among students (Sano et al., 2018). Persistent feedback through wearables reduces users' belief in their own autonomy, with participants monitoring heart rates reporting diminished belief in free will compared to those in control groups (Baker, 2017).

**(c) Behavioral Dependency and Loss of Autonomy**

One of the more insidious psychological impacts of data-driven feedback is the reduction of users' belief in their own autonomy. Wearables that offer continuous performance data, such as heart rate monitors, reinforce reliance on external metrics to guide decision-making (Baker, 2017). This dependency risks diminishing intrinsic motivation and may negatively affect users' belief in free will. Furthermore, the continuous exposure to self-quantification data can create stress by emphasizing outcomes over personal experiences, leading to feelings of inadequacy or failure (Baker, 2017). Real-time social feedback also shapes behavior in unanticipated ways. Users who depend on social media platforms for feedback are more likely to engage in excessive reassurance-seeking, which has been linked to poor mental health outcomes, including depressive symptoms (Nesi & Prinstein, 2015). In professional contexts, constant monitoring of performance can impair creativity and productivity by encouraging individuals to conform to external expectations rather than pursuing innovative solutions (Damian et al., 2018). Adolescents engaging in frequent feedback-seeking behaviors experience higher levels of anxiety and depressive symptoms (Nesi & Prinstein, 2015). Additionally, 419 South Korean participants demonstrated that SNS addiction, driven by social comparison, mediated the relationship between technology overload and reduced well-being (Choi & Lim, 2016).

### (d) Social Dynamics and Communication Patterns

Feedback loops also shape social dynamics, as individuals modify their behavior based on the feedback they receive. This real-time behavioral modification can enhance interpersonal communication in specific settings, such as public speaking or job interviews. However, it may also lead to increased self-consciousness and stress, as individuals become preoccupied with meeting perceived social expectations (Damian et al., 2015). Social feedback systems used in social networks have the potential to enhance social awareness and connectedness. However, they may also create pressures to conform, reducing diversity of thought and behavior. For instance, algorithms that prioritize popular content can lead users to align with majority opinions, fostering echo chambers that diminish cognitive diversity (Wu et al., 2017). In social media, the interplay between technostress and Internet addiction worsens mental health outcomes. One study surveying 1731 participants found positive correlations between distraction-conflict theory behaviors and social media addiction, which adversely impacted workplace productivity (Brooks et al., 2017).

Data-driven feedback loops significantly influence behavior and psychological well-being, offering both opportunities and challenges. While these systems improve motivation, health monitoring, and social connectedness, they also pose risks of technostress, dependency, and diminished autonomy. Striking a balance between leveraging feedback for personal growth and preventing behavioral dependency is crucial. Future research should explore the optimal design of these systems, minimizing negative impacts while promoting user well-being.

### 2.2 Review of Past Works

This section provides a detailed review of past studies on the psychological and behavioral impacts of data-driven feedback loops.

Mert et al. (2023) investigated the impact of social media usage on technostress and cyber incivility among lawyers. The researchers collected survey data from 324 lawyers and analyzed the data using Partial Least Squares Structural Equation Modeling (PLS-SEM). Social media use was identified as a strong predictor of technostress and cyber incivility. Technostress mediated the relationship between social media usage and workplace incivility, indicating that overexposure to social media can impair professional relationships.

Baker (2017) explored the effect of wearable devices on users' belief in free will (BFW). Participants were divided into groups—one group monitored their heart rate using a wearable tracker, another explored the device's features without tracking, and a control group was uninvolved. Result showed that those using wearables to monitor physical states exhibited a decline in BFW, showing that constant monitoring reduced their sense of autonomy. Sano et al. (2018) examined how wearable sensors, and mobile phones can predict self-reported stress and mental health among students. The SNAPSHOT study followed 201 college students for one month, collecting 145,000 hours of wearable sensor data (tracking physical activity and physiology) along with e-diary entries. Machine learning models were used for analysis. Wearable sensor data reached 78.3% accuracy in identifying stress levels and 87% accuracy in predicting mental health status. Modifiable behaviors such as screen time and mobility patterns also contributed to these predictions.

Ostic et al. (2021) analyzed the psychological well-being of social media users by measuring social capital, social isolation, and smartphone addiction. The authors conducted a survey with 940 participants and applied Structural Equation Modeling (SEM) to evaluate direct and indirect effects. Their result showed that social media use positively impacted psychological well-being through bonding and bridging social capital. However, excessive use was associated with addictive behaviors and increased social isolation

Majumder et al. (2020) developed an embedded system integrating wearable devices and social media data to monitor users' mental health. The system collects physiological, behavioral, and social data streams, which are analyzed using a machine-learning algorithm to predict emotional states. The system achieved an emotional state prediction accuracy of 92.9%, demonstrating the effectiveness of real-time feedback in improving mental health outcomes.

McCrory et al. (2020) investigated the psychological impact of highly visual social media platforms like Instagram and Snapchat on adolescents' mental health. They conducted a scoping review of 239 articles, ultimately narrowing down to 25 studies for analysis. The findings revealed mixed evidence regarding the impact of visual social media on adolescent mental health, highlighting the need for more qualitative studies to clarify the psychological effects of such platforms.

Boratto & Vargiu (2020) explored the use of behavioral modeling with wearables, activity trackers, and social media to analyze user behavior.The authors utilized data mining techniques to extract behavioral patterns from sensor-based and social media data, transforming them into actionable insights for feedback systems. The study demonstrated that data-driven feedback loops can effectively provide personalized recommendations and behavioral insights, facilitating behavior change

**2.3 Contribution to Knowledge**

This study fills a critical gap by exploring the impact of wearable devices, tracking apps, and social media on behavioral changes, motivation, and mental well-being within both personal and social contexts. While previous studies, such as those by Baker (2017) and Ostic et al. (2021), focused on individual domains either the impact of wearables on autonomy or the role of social media in fostering social capital this study provides a holistic view of how these technologies intersect and influence multiple dimensions of users' lives. It addresses the underexplored interplay between real-time feedback, social comparison, and psychological well-being, bridging the gap between technology-driven motivation and the risks of technostress and dependency.

The study contributes to knowledge by identifying patterns of behavioral dependency, showing how over-reliance on feedback systems not only shapes individual routines but also influences social dynamics and decision-making. It highlights the dual-edged nature of feedback mechanisms promoting goal achievement while undermining autonomy and mental health hereby extending insights from earlier studies (Mert et al., 2023; Nesi & Prinstein, 2015). This research also emphasizes the impact of feedback-seeking behaviors on productivity, contributing novel insights into how constant engagement with notifications impairs focus and well-being. By integrating these findings, the study offers a comprehensive framework for understanding the balance between the benefits and drawbacks of technology use in everyday life.

**3.0 Methodology**

This study adopts a descriptive survey design. Data from 200 purposefully selected users across health, social, and lifestyle apps was analyzed using surveys (google forms) assessing behavioral changes, motivation, and mental well-being.

**4.0 Results**

**Table 1: Behavioral Changes, Motivation, Mental Well-Being, Social Dynamics and Communication Patterns**

| | Items | SA (%) | A (%) | D (%) | SD (%) |
|---|---|---|---|---|---|
| **Behavioral Changes** | | | | | |
| 1. | I rely on wearable devices or apps to make decisions about my health and fitness (e.g., step count, heart rate monitoring). | 57 (22.7) | 180 (72.1) | 13 (5.2) | 0 (0) |
| 2. | Tracking progress through apps encourages me to prioritize data-driven goals over intrinsic enjoyment | 178 (71.3) | 54 (21.7) | 18 (7.0) | 0 (0) |
| 3. | Frequent use of self-monitoring apps has improved my ability to manage daily routines. | 88 (35) | 150 (60) | 12 (5) | 0 (0) |
| 4 | I feel that my behavior is being shaped by the feedback I receive from wearable or social apps, even when I don't consciously intend it. | 62 (25) | 175 (70) | 12 (5) | 0 (0) |
| 5 | Real-time feedback from wearable devices or apps has changed how I interact in social or professional settings (e.g., adjusting speech or posture). | 0 (0) | 8 (3.4) | 204 (81.4) | 38 (15.2) |
| **Motivation and Mental Well-Being** | | | | | |
| 6. | Positive reinforcement from apps (e.g., badges, streaks) motivates me to maintain healthy behaviors. | 178 (71.3) | 54 (21.7) | 18 (7.0) | 0 (0) |
| 7. | Overuse of social media apps increases my | 19 (7.7) | 188 (75) | 43 (17.3) | 0 (0) |
| | anxiety and stress levels, especially through comparison with others. | | | | |
| 8. | Notifications and reminders from apps sometimes feel overwhelming and increase mental fatigue. | 19 (7.4) | 168 (67) | 58 (23.1) | 6 (2.5) |
| 9. | I feel less autonomous when apps guide my daily decisions, reducing my reliance on personal intuition | 34 (13.4) | 196 (78.5) | 20 (8.1) | 0 (0) |
| 10. | Feedback from wearables or apps has positively impacted my mental well-being by helping me stay on track with my goals. | 68 (27.3) | 149 (59.6) | 32 (12.8) | 1 (0.3) |
| **Social Dynamics and Communication Patterns** | | | | | |
| 11 | I often modify my opinions or actions to align with popular trends or feedback I see on social media | 0 (0) | 123 (49.1) | 91 (36.3) | 36 (14.6) |
| 12 | I feel pressure to receive validation from others through social media interactions (e.g., likes, comments) | 33 (13.2) | 110 (44) | 71 (28.5) | 36 (14.3) |
| 13. | Frequent feedback on social media has increased my self-consciousness in personal and professional settings. | 60 (23.8) | 188 (75.4) | 2 (0.8) | 0 (0) |
| 14. | Feedback-seeking behaviors (e.g., checking notifications frequently) have negatively affected my productivity and well-being. | 88 (35) | 150 (60) | 12 (5) | 0 (0) |

*Strongly Agree (SA) =4, Agree (A) = 3, Disagree (D) = 2, Strongly Disagree (SD) = 1, St.D = Standard Deviation, M= Mean*

Table 1 reveals insights into how wearable devices, tracking apps, and social media influence behavioral changes, motivation, and mental well-being. A majority of the respondents rely on wearable technology for health-related decisions, with 94.8% either agreeing or strongly agreeing that these tools guide their fitness activities. This reliance reflects a shift towards data-driven decision-making. Additionally, the majority of participants (93%) report that tracking progress through apps encourages them to prioritize measurable goals over intrinsic enjoyment, indicating a growing focus on external metrics as motivational drivers.

The feedback provided by wearable devices plays a pervasive role in shaping behavior, with 95% of respondents acknowledging that their actions are unconsciously influenced by these tools, even when they do not intend it. Despite this, real-time feedback from apps has minimal influence on social or professional interactions, as 81.4% disagree that such feedback affects their engagement. This suggests that while technology significantly impacts personal routines, its influence on interpersonal behavior remains limited.

In terms of motivation and mental well-being, digital tools provide both benefits and challenges. Positive reinforcement through apps, such as badges and streaks, serves as a powerful motivator, with 93% of respondents maintaining healthy behaviors due to these incentives. However, the overuse of social media platforms introduces challenges, as 75% of participants report heightened anxiety and stress due to constant comparisons with others. Further, notifications and reminders, which aim to keep users engaged, contribute to mental fatigue, with 74.4% feeling overwhelmed by these prompts. The data further highlights a concern about autonomy, as 78.5% of participants feel less independent when apps dictate their daily decisions, underscoring the potential erosion of personal intuition.

In terms of social dynamics and communication patterns, the findings also shed light on the impact of social feedback on self-perception and productivity. A significant number of respondents experience pressure to seek validation through social media, with 57.2% admitting they feel compelled to gain likes and comments. Frequent feedback from social media amplifies self-consciousness, affecting both personal and professional interactions for the majority of respondents. This suggests that social interactions are increasingly shaped by online feedback, making individuals more mindful of how they present themselves digitally. Furthermore, feedback-seeking behavior negatively impacts productivity and well-being, as 95% of respondents acknowledge that constantly checking notifications detracts from their focus.

**4.1 Discussions**

The findings from this study shows that data-driven feedback loops in wearables, social media, and mobile apps significantly influence behavior, motivation, and mental well-being. Wearable devices, by providing real-time feedback, foster positive behavioral changes by increasing self-awareness and accountability. This corroborates research by Baker (2017), who found that users of health and fitness wearables experience improved motivation and goal achievement through continuous tracking of progress. Similarly, Ostic et al. (2021) identified that social media use fosters psychological well-being through bonding and bridging social capital, creating a sense of belonging and connectedness. This suggests that these technologies can promote social and personal development by enhancing motivation and social engagement.

However, the study also reflects the potential negative consequences of excessive reliance on digital feedback, including mental fatigue, stress, and reduced autonomy. Hsiao et al. (2017) documented that compulsive use of social media apps increases anxiety, mental fatigue, and stress, corroborating the results, where 75% of respondents experienced heightened anxiety from social comparisons. Additionally, the study by Mert et al. (2021) reinforces these outcomes by showing how constant notifications and online engagement contribute to technostress and workplace incivility.

The findings also underscore how real-time feedback from apps shapes self-perception and social interactions. Users become more self-conscious and feel pressured to conform to social norms, aligning with findings by Nesi and Prinstein (2015), who identified that social comparison on digital platforms triggers anxiety and depressive symptoms, particularly among adolescents. Similarly, Damian et al. (2015) demonstrated that feedback loops during public speaking enhance communication but may lead to performance pressure, aligning with participants' feelings of reduced autonomy reported in the survey.

Further, the findings suggests that constant feedback-seeking behavior can negatively impact productivity and decision-making. This aligns with Sano et al. (2018), who found that students' screen time and behavior patterns were predictive of stress and mental health issues, reinforcing the idea that dependency on technology for decision-making diminishes intrinsic motivation and well-being. Baker (2017) also found that individuals tracking biometric data using wearables showed a decline in belief in free will, reflecting the erosion of autonomy discussed in the findings. While these systems promote healthy behaviors and social connections, they also pose challenges such as technostress, dependency, and social comparison. Balancing the benefits and drawbacks of feedback mechanisms is essential, as suggested by Wu et al. (2017), who warn of the risks of algorithmic influence fostering echo chambers and reducing cognitive diversity.

Further, AI models can be designed to detect patterns indicating user stress, fatigue, or negative emotional states. Parekh et al. (2020) reported that artificial intelligence is one of the essential tactics to detect or monitor fatigue. The authors used Artificial neural network, wavelet transform, data analysis of mouse interaction and keyboard patterns, image analysis, kernel learning algorithms, relation of fatigue and anxiety, and heart rate data examination studies to precisely assess the source factors and features which influenced the recognition of fatigue (Parekh et al., 2020). By using adaptive algorithms, such systems could automatically adjust the frequency and tone of notifications to prevent cognitive overload. This would help users maintain productivity and well-being without succumbing to the burnout often induced by continuous feedback. Also, AI systems can enhance the user experience by minimizing social comparison. By personalizing performance metrics to individual goals rather than comparative metrics, AI models can promote healthier engagement and reduce the adverse emotional impacts of social media validation systems. Zhu et al., 2020 used multi-armed bandits AI technique to assess 53 users using m health app for physical activity. Their result indicates that there is some evidence that motivation can be increased using the AI-based personalization of social comparison. For example, dynamic goal setting and AI-curated feedback can motivate users based on their progress, rather than through competition with others.

## 5.0 Conclusion

The findings revealed that while wearable devices, tracking apps, and social media provide valuable tools for enhancing self-awareness, motivation, and social connectedness, they also introduce challenges such as technostress, dependency, and reduced autonomy. Positive reinforcement through these technologies encourages goal achievement, but the overemphasis on external validation and data-driven behavior risks diminishing intrinsic motivation and increasing anxiety. Real-time feedback plays a significant role in shaping behavior and social interactions, yet it also fosters self-consciousness and impacts productivity by encouraging constant feedback-seeking behavior.

## 6.0 Recommendation

To maximize the benefits while minimizing negative outcomes, users should adopt a balanced approach to technology use, setting boundaries to prevent mental fatigue and over-reliance on feedback systems. Developers of wearable and social media technologies should focus on optimizing notification systems to reduce cognitive overload and foster healthier engagement. Institutions and workplaces can implement digital literacy programs to promote mindful usage and reinforce the importance of intrinsic motivation. Developers should also adopt best practices that balance motivation and mental well-being. Incorporating AI-driven feedback mechanisms will allow systems to dynamically respond to user fatigue and stress, tailoring notifications to prevent burnout. Mobile app developers can integrate adaptive systems into product design by:

i. Developing stress-aware features that modify notifications based on real-time behavior patterns.
ii. Promoting intrinsic motivation by replacing competitive metrics with individualized goal tracking.
iii. Implementing mindful notification strategies that encourage user engagement without triggering mental fatigue.

## 7.0 Future Research

i. Future studies could examine how different AI algorithms, such as reinforcement learning, impact user well-being by personalizing feedback in real-time.
ii. Research could explore whether adaptive AI systems promote sustainable behavior change or inadvertently increase dependence on external feedback.
iii. Further studies can focus on how AI-based feedback systems affect vulnerable groups, such as adolescents or individuals with mental health conditions.
iv. Future research could investigate the use of AI-enhanced feedback systems within specific industries such as health, education, or workplace settings.
v. Further study can be done on the ethical considerations surrounding AI feedback, particularly regarding data privacy, transparency, and user autonomy.

**References**

Baker, D. (2017). Wearables and User Interface Design: Impacts on Belief in Free Will. , 210-217. https://doi.org/10.1007/978-3-319-58750-9_30.

Boratto, L., & Vargiu, E. (2020). Data-driven user behavioral modeling: from real-world behavior to knowledge, algorithms, and systems. *Journal of Intelligent Information Systems*, 54, 1 - 4. https://doi.org/10.1007/s10844-020-00593-x.

Brooks, S., Longstreet, P., & Califf, C. (2017). Social media Induced Technostress and its Impact on Internet Addiction: A Distraction-conflict Theory Perspective. *AIS Trans. Hum. Comput. Interact.*, 9, 2. https://doi.org/10.17705/1THCI.00091.

Choi, S., & Lim, M. (2016). Effects of social and technology overload on psychological well-being in young South Korean adults: The mediatory role of social network service addiction. *Comput. Hum. Behav.*, 61, 245-254. https://doi.org/10.1016/j.chb.2016.03.032.

Collins, D., Harvey, S., Lavender, I., Glozier, N., Christensen, H., & Deady, M. (2020). A Pilot Evaluation of a Smartphone Application for Workplace Depression. *International Journal of Environmental Research and Public Health*, 17. https://doi.org/10.3390/ijerph17186753.

Damian, I., Baur, T., & André, E. (2016). Measuring the impact of multimodal behavioural feedback loops on social interactions. *Proceedings of the 18th ACM International Conference on Multimodal Interaction*. https://doi.org/10.1145/2993148.2993174.

Damian, I., Tan, C., Baur, T., Schöning, J., Luyten, K., & André, E. (2015). Augmenting Social Interactions: Realtime Behavioural Feedback using Social Signal Processing Techniques. *Proceedings of the 33rd Annual ACM Conference on Human Factors in Computing Systems*. https://doi.org/10.1145/2702123.2702314.

Hadi, R., & Valenzuela, A. (2019). Good Vibrations: Consumer Responses to Technology-Mediated Haptic Feedback. *Journal of Consumer Research*. https://doi.org/10.1093/jcr/ucz039.

Hsiao, K., Shu, V., & Huang, T. (2017). Exploring the effect of compulsive social app usage on technostress and academic performance: Perspectives from personality traits. *Telematics Informatics*, 34, 679-690. https://doi.org/10.1016/j.tele.2016.11.001.

Kaur, J., Lavuri, R., Parida, R., & Singh, S. (2023). Exploring the Impact of Gamification Elements in Brand Apps on the Purchase Intention of Consumers. *J. Glob. Inf. Manag.*, 31, 1-30. https://doi.org/10.4018/jgim.317216.

Lee, H., Jamieson, J., Reis, H., Beevers, C., Josephs, R., Mullarkey, M., O'Brien, J., & Yeager, D. (2020). Getting Fewer "Likes" Than Others on Social Media Elicits Emotional Distress Among Victimized Adolescents. *Child development*. https://doi.org/10.1111/cdev.13422.

Liu, Z., Lin, X., Wang, X., & Wang, T. (2020). Self-Regulation Deficiency in Predicting Problematic Use of Mobile Social Networking Apps: The Role of Media Dependency. *Decis. Sci.*, 53, 827-855. https://doi.org/10.1111/deci.12495.

Majumder, A., Dedmondt, J., Jones, S., & Asif, A. (2020). A Smart Cyber-Human System to Support Mental Well-Being through Social Engagement. *2020 IEEE 44th Annual Computers, Software, and Applications Conference (COMPSAC)*, 1050-1058. https://doi.org/10.1109/COMPSAC48688.2020.0-134.

Matthes, J., Karsay, K., Schmuck, D., & Stevic, A. (2020). "Too much to handle": Impact of mobile social networking sites on information overload, depressive symptoms, and well-being. *Comput. Hum. Behav.*, 105, 106217. https://doi.org/10.1016/j.chb.2019.106217.

Maza, M., Fox, K., Kwon, S., Flannery, J., Lindquist, K., Prinstein, M., & Telzer, E. (2023). Association of Habitual Checking Behaviors on Social Media With Longitudinal Functional Brain Development.. *JAMA pediatrics*. https://doi.org/10.1001/jamapediatrics.2022.4924.

McCrory, A., Best, P., & Maddock, A. (2020). The relationship between highly visual social media and young people's mental health: A scoping review. *Children and Youth Services Review*. https://doi.org/10.1016/j.childyouth.2020.105053.

Mert, I., Şen, C., & Abubakar, A. (2023). Impact of social Media usage on technostress and cyber incivility. *Information Development*. https://doi.org/10.1177/02666669231204954.

Nesi, J., & Prinstein, M. (2015). Using Social Media for Social Comparison and Feedback-Seeking: Gender and Popularity Moderate Associations with Depressive Symptoms. *Journal of Abnormal Child Psychology*, 43, 1427 - 1438. https://doi.org/10.1007/s10802-015-0020-0.

Oberst, U., Wegmann, E., Stodt, B., Brand, M., & Chamarro, A. (2017). Negative consequences from heavy social networking in adolescents: The mediating role of fear of missing out.. *Journal of adolescence*, 55, 51-60. https://doi.org/10.1016/j.adolescence.2016.12.008.

Ostic, D., Qalati, S., Barbosa, B., Shah, S., Vela, E., Herzallah, A., & Liu, F. (2021). Effects of Social Media Use on Psychological Well-Being: A Mediated Model. *Frontiers in Psychology*, 12. https://doi.org/10.3389/fpsyg.2021.678766.

Parekh, V., Shah, D., & Shah, M. (2020). Fatigue detection using artificial intelligence framework. *Augmented Human Research*, *5*(1), 5.

Sano, A., Taylor, S., McHill, A., Phillips, A., Barger, L., Klerman, E., & Picard, R. (2018). Identifying Objective Physiological Markers and Modifiable Behaviors for Self-Reported Stress and Mental Health Status Using Wearable Sensors and Mobile Phones: Observational Study. *Journal of Medical Internet Research*, 20. https://doi.org/10.2196/jmir.9410.

Small, G., Lee, J., Kaufman, A., Jalil, J., Siddarth, P., Gaddipati, H., Moody, T., & Bookheimer, S. (2020). Brain health consequences of digital technology use. *Dialogues in Clinical Neuroscience*, 22, 179 - 187. https://doi.org/10.31887/DCNS.2020.22.2/gsmall.

Takeuchi, H., Suwa, K., Kishi, A., Nakamura, T., Yoshiuchi, K., & Yamamoto, Y. (2022). The Effects of Objective Push-Type Sleep Feedback on Habitual Sleep Behavior and Momentary Symptoms in Daily Life: mHealth Intervention Trial Using a Health Care Internet of Things System. *JMIR mHealth and uHealth*, 10. https://doi.org/10.2196/39150.

Wang, Y., Weber, I., & Mitra, P. (2016). Quantified Self Meets Social Media: Sharing of Weight Updates on Twitter. *Proceedings of the 6th International Conference on Digital Health Conference*. https://doi.org/10.1145/2896338.2896363.

Wolf, T., Jahn, S., Hammerschmidt, M., & Weiger, W. (2020). Competition versus cooperation: How technology-facilitated social interdependence initiates the self-improvement chain. *International Journal of Research in Marketing*. https://doi.org/10.1016/j.ijresmar.2020.06.001.

Wu, C., Zhang, Y., Jia, J., & Zhu, W. (2017). Mobile Contextual Recommender System for Online Social Media. *IEEE Transactions on Mobile Computing*, 16, 3403-3416. https://doi.org/10.1109/TMC.2017.2694830.

Zhao, J., Freeman, B., & Li, M. (2016). Can Mobile Phone Apps Influence People's Health Behavior Change? An Evidence Review. *Journal of Medical Internet Research*, 18. https://doi.org/10.2196/jmir.5692.

Zhu, J., Dallal, D. H., Gray, R. C., Villareale, J., Ontañón, S., Forman, E. M., & Arigo, D. (2021). Personalization paradox in behavior change apps: lessons from a social comparison-based personalized app for physical activity. *Proceedings of the ACM on Human-Computer Interaction*, *5*(CSCW1), 1-21.